\documentclass[12pt,epsfig,color]{article}
\usepackage{amsmath}
\usepackage{epsfig}
\usepackage{amssymb}
\input{epsf}
\newcommand{\mysection}{\setcounter{equation}{0}\section}

\def\beq{\begin{equation}}
\def\eeq{\end{equation}}
\def\beqa{\begin{eqnarray}}
\def\eeqa{\end{eqnarray}}

\newcommand\as{\alpha_{\mathrm{S}}}
\newcommand\f[2]{\frac{#1}{#2}}

\def\beq{\begin{equation}}
\def\eeq{\end{equation}}
\def\beeq{\begin{eqnarray}}
\def\eeeq{\end{eqnarray}}
\def\to{\rightarrow}
\def\nn{\nonumber}

\def\b0{b_0}

\begin{document}

\begin{titlepage}
\renewcommand{\thefootnote}{\fnsymbol{footnote}}
\begin{flushright}
BNL-NT-07/55 \\
hep-ph/???
     \end{flushright}
\par \vspace{10mm}
\begin{center}
{\large \bf
Single-inclusive hadron production in transversely polarized $pp$  \\[4mm]
 and $\bar{p}p$ collisions with threshold resummation}

\end{center}
\par \vspace{2mm}
\begin{center}
{\bf Daniel de Florian${}^{\,a}$,}
\hskip .2cm
{\bf Werner Vogelsang${}^{\,b}$,}
\hskip .2cm
and
\hskip .2cm
{\bf Federico Wagner${}^{\,a}$  }\\
\vspace{5mm}
${}^{a}\,$Departamento de F\'\i sica, FCEYN, Universidad de Buenos Aires,\\
(1428) Pabell\'on 1 Ciudad Universitaria, Capital Federal, Argentina\\
${}^{b}\,$Physics Department, Brookhaven National Laboratory, 
Upton, NY 11973, U.S.A.\\

\end{center}


\par \vspace{9mm}
\begin{center} {\large \bf Abstract} \end{center}
\begin{quote}
\pretolerance 10000
We investigate the resummation of large logarithmic perturbative corrections 
to the partonic cross sections for single-inclusive high-$p_T$ hadron 
production in collisions of transversely polarized hadrons. 
We perform the resummation to next-to-leading logarithmic accuracy. 
Phenomenological results are given for $\bar{p}p$ collisions at 
center-of-mass energy $\sqrt{S}=14.5$ GeV and for $pp$ collisions at 
$\sqrt{S}=62.4$ GeV and at $\sqrt{S}=10$ GeV, which are relevant for 
possible experiments at the GSI-FAIR, RHIC and J-PARC facilities, 
respectively. We find significant enhancements of the spin-dependent 
and spin-averaged cross sections, but a decrease
of the double-spin asymmetry $A^{\pi}_{TT}$.
\end{quote}

\end{titlepage}

\setcounter{footnote}{1}
\renewcommand{\thefootnote}{\fnsymbol{footnote}}


\section{Introduction}
In spite of extensive studies in recent years, the partonic structure of 
spin-$1/2$ nucleons is not yet completely known. Among the leading-twist 
collinear parton distributions, the unpolarized ($f$), longitudinally
polarized ($\Delta f$), and transversely polarized ($\delta f$) densities,
there is so far very little information about the latter. 
These ``transversity'' distributions $\delta f$ are 
defined~\cite{Jaffe,Ralston,Artru} as the differences of probabilities 
to find a parton of flavor $f$ at scale $\mu$ and light-cone momentum 
fraction $x$ with its spin aligned ($\uparrow \uparrow$) or 
anti-aligned ($\uparrow \downarrow$) with that of the transversely 
polarized nucleon:
\begin{equation}\label{eq:ftrans}
\delta f(x,\mu)\equiv f_{\uparrow \uparrow}(x,\mu) - f_{\uparrow \downarrow}
(x,\mu) \; .
\end{equation}
As is well-known, there is no leading-twist gluon transversity distribution,
due to the odd chirality of transversity and angular momentum conservation. 
The unpolarized parton distributions are recovered by taking the sum in 
Eq.~(\ref{eq:ftrans}). 

Unlike the longitudinally polarized distribution functions which can be 
measured directly in deep-inelastic scattering (DIS), transversity is not 
accessible in inclusive DIS due to its chiral-odd nature~\cite{Ralston}. 
Only recently has a very first ``glimpse'' of transversity been 
obtained from a combined analysis~\cite{Anselmino1} of data for 
single-transverse spin asymmetries in semi-inclusive deep-inelastic 
scattering (SIDIS)~\cite{hermes} and $e^+e^-$ annihilation~\cite{belle}. 
This analysis relies on the extraction 
of the Collins functions~\cite{Collins} from $e^+e^-$ data, which then 
give access to transversity in SIDIS spin asymmetries. Apart from
further pursuing such studies, it will be highly desirable in the future
to have more direct probes of transversity. These are available in
double-transverse spin asymmetries in hadronic collisions, 
\begin{equation}\label{eq:Att}
A_{TT}\equiv 
\frac{\frac{1}{2}[d\sigma(\uparrow \uparrow) - 
d\sigma(\uparrow \downarrow)]}{\frac{1}{2}[d\sigma(\uparrow \uparrow) + 
d\sigma(\uparrow \downarrow)]}\equiv \frac{d\delta \sigma}{ d  \sigma}\; ,
\end{equation}
where the arrows denote the transverse polarization of the scattering
hadrons. A program of polarized $pp$ collisions is now well underway
at the BNL Relativistic Heavy Ion Collider (RHIC), and measurements
of $A_{TT}$ for various reactions should be feasible with sufficient
beam-time for transverse polarization~\cite{spinplan}. In the more 
distant future, there is also hope to have transversely polarized 
$\bar{p}p$ collisions at the GSI-FAIR facility, where there are plans 
to have an asymmetric polarized $\bar{p}p$ collider~\cite{ref:gsi-fair}. 
Likewise, transversely polarized $pp$ collisions could become a possibility
at the J-PARC facility~\cite{ref:jparc}. In view of these opportunities,
it is important to supply a theoretical framework that adequately 
describes the processes of interest, allowing a reliable extraction of
transversity from hopefully forthcoming data.

In the present paper, we will focus on single-inclusive hadron production
in transversely polarized hadronic collisions, $p^\uparrow p^\uparrow
\to hX$, $\bar{p}^\uparrow p^\uparrow \to hX$, where the hadron $h$ will 
for our purposes be a pion and has large transverse momentum $p_T$. 
Compared to the Drell-Yan process, which
is usually considered the ``golden channel'' for transversity measurements
in hadronic 
scattering~\cite{Ralston,Cortes,pdfmodel,Ji,Vogelsang2,Shim,Anselmino,Ratcliffe},
the spin 
asymmetry for hadron production is typically considerably smaller, mostly 
due to the large contribution from gluonic scattering in the denominator 
of the asymmetry~\cite{Artru,Ji,Jaffe2}. On the other hand, pions are very 
copiously produced in hadronic scattering, resulting in much smaller 
statistical uncertainties, and $A_{TT}$ in single-inclusive hadron 
production could thus conceivably present a viable alternative to 
Drell-Yan at RHIC as well as at GSI-FAIR and J-PARC~\cite{Muk}. 

Theoretical calculations of high-$p_T$ pion production are based
on the factorization theorem~\cite{FT}, which states that the cross 
section may be factorized in terms of collinear convolutions of universal
parton distribution for the initial hadrons (in the spin-dependent
case, transversity), a fragmentation function for the final-state 
pion, and short distance parts that describe the hard interactions 
of the partons and are amenable to QCD perturbation theory. The latter
thus have an expansion in the strong coupling $\as$, starting
with a leading-order (LO) term, followed by a next-to-leading order (NLO)
correction, etc. For the process we are interested in here, it 
was found that at RHIC energies theoretical NLO calculations 
are very successful in describing the experimental 
data~\cite{cross_phenix,cross_star,cross_brahms}. 
The NLO corrections relevant for
the process $pp\to\pi X$ with transversely polarized protons are
available in~\cite{Muk}. 

For much lower energies, as typically available in fixed-target 
experiments, our previous work~\cite{DW1,DWF} has shown that
the NLO framework is no longer sufficient but that all-order 
resummations of large logarithmic corrections are needed.  
Here the value of $x_T\equiv 2 p_T/\sqrt{S}$, with $\sqrt{S}$ 
the center-of-mass (c.m.) energy of the collision, is generally 
quite large, $x_T\gtrsim 0.1$. It turns out that the partonic 
hard-scattering cross sections relevant for $pp \to \pi X$ 
are then largely probed in the ``threshold''-regime, 
where the initial partons have just enough energy to produce the
high-transverse momentum parton that subsequently fragments into the
hadron, and its recoiling counterpart. Relatively little phase space 
is then available for additional radiation of partons. In particular,
gluon radiation is inhibited and mostly constrained to the emission
of soft and/or collinear gluons. The cancellation of infrared singularities 
between real and virtual diagrams then leaves behind large double- and
single-logarithmic corrections to the partonic cross sections. These
logarithms appear for the first time at NLO, where they arise as terms 
of the form 
$\as\ln^2(1-\hat{x}_T^2)$ in the rapidity-integrated cross section,
where $\hat{x}_T\equiv 2 \hat{p}_T/\sqrt{\hat{s}}$ with $\hat{p}_T$
the transverse momentum of the produced parton and $\hat{s}$ the
c.m. energy of the initial partons. At yet higher ($k$th) order
of perturbation theory, the double-logarithms are of the form 
$\as^k\ln^{2k}(1-\hat{x}_T^2)$. When the threshold regime dominates,
it is essential to take into account the large logarithms to all 
orders in the strong coupling $\as$, a technique known as ``threshold
resummation''~\cite{resumm,KOS}. In our earlier work, we examined the
effects of threshold resummation on the single-inclusive
hadron cross section in~\cite{DW1} and on the double-longitudinal 
spin asymmetry $A_{LL}$~\cite{DWF}. We found very significant
enhancements of the cross section by resummation, which in fact lead 
to a relatively good agreement between resummed theory and experimental 
data. We also found a moderate decrease of the resummed $A_{LL}$, 
with respect to NLO. We concluded that threshold resummation 
is an essential part of the theoretical description in the
fixed-target kinematic regime. Its effects at higher energies (such as
at RHIC) are much smaller, even though it has to be said that one is 
typically much further away from the threshold regime here, so that 
the applicability of threshold resummation is not entirely clear. 

In the present paper, we adapt threshold resummation to the 
case of transverse polarization of the scattering partons, which 
is relevant for studies of $A_{TT}$. From a technical point of view,
this is relatively straightforward, given our previous work~\cite{DW1,DWF}. 
We will also use our results to obtain phenomenological predictions 
for cross sections and the double-transverse spin asymmetries in
the kinematic regimes to be accessed by the possible GSI-FAIR, RHIC and
J-PARC measurements. The remainder of this paper is organized as follows. 
In Sec.~2, the general framework for the resummed single-inclusive 
hadron cross section is briefly reviewed, and the ingredients that 
are specific to transverse polarization are provided. Section~3 presents 
phenomenological results. We conclude in Sec.~4, and two Appendices collect
some relevant expressions for each of the partonic subprocesses.

\section{The $p_T$ differential cross section in perturbation theory 
\label{sec2}}

We will for simplicity consider the cross section integrated over 
all rapidities of the produced hadron $h$, which turns out to simplify 
the analysis significantly~\cite{DW1,DWF}. The factorized spin-dependent 
cross section differential in the hadron's transverse momentum $p_T$ and 
its azimuthal angle $\phi$ with respect to the initial transverse 
spin directions can then be written as~\cite{DW1}
\begin{align}
\label{eq:crosssect} \f{p_T^3\, d\delta\sigma(x_T)}{dp_T d\phi} = 
\sum_{a,b,c}\, &
\int_0^1 dx_1 \, \delta f_{a}\left(x_1,\mu^2\right) \, \int_0^1
dx_2 \, \delta f_{b}\left(x_2,\mu^2\right) \, \int_0^1 dz
\,z^2\, D_{h/c}\left(z,\mu^2\right) \, \nn \\ &\int_0^1
d\hat{x}_T \, \, \delta\left(\hat{x}_T-\f{x_T}{z\sqrt{x_1
x_2}}\right) \, \int_{\hat{\eta}_{-}}^{\hat{\eta}_{+}} d\hat{\eta}
\, \f{\hat{x}_T^4 \,\hat{s}}{2} \,
 \f{d\delta \hat{\sigma}_{ab\rightarrow cX}(\hat{x}_T^2,\hat{\eta},
\as(\mu),\mu)}{d\hat{x}_T^2 d\hat{\eta} d\phi} \, ,
\end{align}
where the $\delta f_{a,b}$ are the transversity parton distribution 
functions defined in Eq.~(\ref{eq:ftrans}), and where the $D_{h/c}$ are 
the parton-to-hadron fragmentation functions. Long- and short-distance 
contributions are separated by a factorization scale $\mu$. We take this 
scale to be the same as the renormalization scale in the strong coupling 
constant. As before, $x_T \equiv 2 p_T/ \sqrt{S} $, and its partonic 
counterpart is $\hat{x}_T$. We have $\hat{\eta}_{+}=-\hat{\eta}_{-}=
\ln\left[(1+\sqrt{1-\hat{x}_T^2})/\hat{x}_T\right]$. 
The sum in Eq.~(\ref{eq:crosssect}) runs 
over all partonic subprocesses $ab\to cX$, with partonic cross 
sections $d\delta \hat{\sigma}_{ab\rightarrow cX}$, defined 
similarly to the numerator of Eq.~(\ref{eq:Att}) for transversely 
polarized initial
partons. As we have mentioned before, for the transversity case there are
no subprocesses which initial gluons. We therefore only have four 
subprocesses that contribute~\cite{Muk}:
\begin{equation}
 q q \to q X, \quad  q \bar{q} \to q X, \quad  q \bar{q} \to q' X, 
\quad  q \bar{q} \to g X. \nonumber
\end{equation}
The cross sections for these have a well-known characteristic dependence 
on the azimuthal angle $\phi$. In the c.m frame, taking the momentum and spin 
directions of the initial hadrons as the $z$- and $x$-axes, respectively,
this dependence is of the form $\cos(2\phi)$. We note that 
the expression for the spin-averaged cross section is identical
to that in Eq.~(\ref{eq:crosssect}), with the transversity distributions
and polarized partonic cross sections replaced by their standard spin-averaged
counterparts. Here the sum of course also runs over partonic channels 
with gluons in the initial state as well. 

We will only briefly describe the technical aspects of the resummation
of the threshold logarithms, pointing out the specifics of the transversity
case. All other details may be found in our previous papers~\cite{DW1,DWF}. 
The resummation of the soft gluon contributions is achieved by taking 
a Mellin transform of the cross section in the scaling variable $x_T^2$:
\begin{align}
\label{eq:moments}
\frac{d\delta \sigma(N)}{d \phi}\equiv \int_0^1 dx_T^2 \, \left(x_T^2 
\right)^{N-1} \;
\f{p_T^3\, d\delta \sigma(x_T)}{dp_T d\phi} \, .
\end{align}
In the same way, the partonic cross section can be expressed as
\begin{align}
\label{momdef}
\frac{ d \delta \hat{\sigma}_{ab\to cX}(N)}{d \phi}\equiv \int_0^1 
d\hat{x}_T^2  \, (\hat{x}_T^2)^{N-1}\, \int_{\hat{\eta}_{-}}^{\hat{\eta}_{+}} 
d\hat{\eta} \, \frac{\hat{x}_T^4 \, \hat{s}}{2} \; \frac{d\delta 
\hat{\sigma}_{ab \to cX}(\hat{x}_T^2,\hat{\eta})}{d\hat{x}_T^2 
d\hat{\eta} d\phi} \, ,
\end{align}
where we from now on suppress the scale dependence of the cross section.
In Mellin-moment space the convolutions in Eq.~(\ref{eq:crosssect}) 
become ordinary products, and threshold logarithms appear as logarithms 
in the moment variable $N$. As we discussed in~\cite{DW1}, 
the resummed partonic cross section for each subprocess 
can be written in the rather simple form
\begin{align}
\label{eq:res}
\frac{d \delta \hat{\sigma}^{{\rm (res)}}_{ab\to cd} (N)}{d \phi}= 
\delta C_{ab\to cd}\,
\Delta^a_N\, \Delta^{b}_N\, \Delta^{c}_N\,
J^{d}_N\, \left[ \sum_{I} \delta G^{I}_{ab\to cd}\,
\Delta^{{\rm (int)} ab\rightarrow cd}_{I\, N}\right] \,
\frac{ d\delta \hat{\sigma}^{{\rm (Born)}}_{ab\to cd} (N)}{d\phi} \;  ,
\end{align}
where $\delta \hat{\sigma}^{{\rm (Born)}}_{ab\to cd}(N)$ denotes the 
LO term in the perturbative expansion of Eq.~(\ref{momdef}) for a given
partonic process. Each of the functions  $J^{d}_N$, $\Delta^{i}_N$, 
$\Delta^{{\rm (int)} ab\rightarrow cd}_{I\, N}$ is an exponential and 
embodies part of the resummation. These terms all coincide with the 
corresponding ones for the spin-averaged case and can be found, for example,
in~\cite{DW1}. ${\Delta}^{a,b}_N$ represent the effects 
of soft-gluon radiation collinear to initial partons $a,b$, and
similarly for $\Delta^c_N$ for the final-state fragmenting parton $c$.
The function $J^{d}_N$ embodies collinear, soft or hard, emission by 
the ``non-observed'' recoiling parton $d$. Large-angle soft-gluon 
emission is accounted for by the factors ${\Delta}^{{\rm (int)} 
ab\rightarrow cd}_{I\, N}$, which depend on the color configuration 
$I$ of the participating partons. Each of the ${\Delta}^{{\rm (int)} ab
\rightarrow cd}_{I\, N}$ is given as 
\begin{align}\label{Dintfct}
\ln{\Delta}^{{\rm (int)} ab\rightarrow cd}_{I\, N} &=
\int_0^1 \f{z^{N-1}-1}{1-z} D_{I\, ab\to cd}(\as((1-z)^2 p_T^2))dz \; .
\end{align}
A sum over the color configurations occurs in Eq.~(\ref{eq:res}), 
with $\delta G^{I}_{ab\to cd}$ representing a weight for each $I$, 
such that $\sum_I \delta G^{I}_{ab\to cd}=1$. Finally, the 
coefficients $\delta C_{ab\to cd}$ contain $N-$independent hard 
contributions arising from one-loop virtual corrections. 
Their perturbative expansion reads:
\begin{eqnarray}
\delta C_{ab\to cd} = 1 + \frac{\as}{\pi} \, \delta C_{ab\to cd}^{(1)} + 
{\cal O}(\as^2) \; .
\end{eqnarray}
They can be determined for each partonic channel by expanding the 
resummed cross section in Eq.~(\ref{eq:res}) to first order in $\as$
and comparing to the full analytic NLO calculation of Ref.~\cite{Muk}.

The only differences between the resummation formulas for the 
spin-dependent and the spin-averaged cases reside in the coefficients 
$\delta G^{I}_{ab\to cd}, \delta C_{ab\to cd}$ and of course in 
the Born cross sections. These terms are all related to hard
scattering, which is in general spin-dependent. We collect 
the moment-space expressions for the spin-dependent Born cross 
sections, as well as the $\delta G^{I}_{ab\to cd}$ and the
$\delta C_{ab\to cd}$ for the various subprocess in Appendix~A.
The corresponding expressions for the spin-averaged case may
all be found in our previous paper~\cite{DW1}, to which we also
refer the reader for further details of the calculation of the
coefficients.

In order to obtain a resummed cross section in $x_T^2$ space, we need an 
inverse Mellin transform. We use the {\em Minimal Prescription} proposed 
in Ref.~\cite{Catani:1996yz} to treat the singularity of the perturbative
strong coupling constant in the resummed exponent. In order to make full 
use of the available fixed-order cross section, which in our case
is NLO (${\cal O}(\as^3)$)~\cite{Muk}, we perform a matching to this cross 
section. We expand the resummed cross section to ${\cal O}(\as^3)$, 
subtract the expanded result from the resummed one, and add the full 
NLO cross section:
\begin{align}
\label{hadnres}
\f{p_T^3\, d\delta \sigma^{\rm (match)}(x_T)}{dp_T \, d\phi} &= \sum_{a,b,c}\,
\;\int_{{\mathrm{Min.\, Prescr.}}}
\;\frac{dN}{2\pi i} \;\left( x_T^2 \right)^{-N+1}
\; \delta f_{a}(N,\mu^2) \; \delta f_{b}(N,\mu^2) \;
D_{c/h}(2N+1,\mu^2)
 \nn \\
&\times \left[ \;
\frac{d \delta \hat{\sigma}^{\rm (res)}_{ab\to cd} (N)}{d \phi}
- \left. \frac{d \delta \hat{\sigma}^{{\rm (res)}}_{ab\to cd} (N)}{d \phi}
\right|_{{\cal O}(\as^3)} \, \right]
+\f{p_T^3\, d\delta \sigma^{\rm (NLO)}(x_T)}{dp_T \, d \phi}
 \;\;,
\end{align}
where $\delta \hat{\sigma}^{{\rm (res)}}_{ab\to cd} (N)$ is the 
transversely polarized resummed cross section for the partonic channel 
$ab\to cd$ as given in Eq.~(\ref{eq:res}), and where the Mellin integration
contour is chosen according to the Minimal Prescription (see 
Ref.~\cite{Catani:1996yz}). In this way, NLO is taken into account in 
full, and the soft-gluon contributions beyond NLO are resummed to 
next-to-leading logarithm (NLL). Any double-counting of perturbative 
orders is avoided.


\section{Phenomenological Results \label{sec3}}

We will now apply the threshold resummation formalism outlined above
to make some predictions for cross sections and spin asymmetries
for single-inclusive hadron production in transversely polarized 
scattering. We will consider $\pi^0$ production in $\bar{p}p$ 
collisions at center-of-mass energy $\sqrt{S}=14.5$ GeV, and in
$pp$ collisions at $\sqrt{S}=62.4$ GeV and at $\sqrt{S}=10$ GeV. 
These conditions correspond to the possible experiments at 
GSI-FAIR~\cite{ref:gsi-fair}, RHIC and J-PARC~\cite{ref:jparc}, 
respectively, that we mentioned in the introduction. In the GSI case, 
this energy would be achieved for an asymmetric collider with polarized 
antiprotons of energy $E_{\bar{p}}=15$ GeV colliding with protons 
of energy $E_p=3.5$ GeV, while for J-PARC a fixed-target set-up is 
envisaged. We note that the current ``default'' energy of the RHIC 
polarized $pp$ collider is $\sqrt{S}=200$~GeV, and only a brief run at 
$\sqrt{S}=62.4$ GeV has been performed. However, given that at high
energies the partonic threshold regime makes a less dominant contribution
to the cross sections, we refrain from presenting
results for $\sqrt{S}=200$~GeV. As we have shown in Ref.~\cite{DWF}, 
at $\sqrt{S}=62.4$ GeV the threshold approximation is still 
reasonably good, and in fact resummation leads to a visible
improvement between the theoretical description and the RHIC 
data reported in~\cite{TAN}. In any case, it is not yet decided 
at which energy measurements of $A_{TT}$ will predominantly
be performed at RHIC. 

We first need to choose sets of parton distribution and pion 
fragmentation functions. For the spin-averaged distributions 
we use the CTEQ6M~\cite{CTEQ} set. We follow Ref.~\cite{pdfmodel}
to model the essentially unknown transversity distributions
by saturating the Soffer inequality~\cite{Soffer} at a low
initial scale $Q_0\sim 0.6$~GeV for the 
evolution (for details see Ref.~\cite{pdfmodel}). 
Finally, for the pion fragmentation functions we choose the 
``de Florian-Sassot-Stratmann'' set~\cite{fDSS}.
According to Eq.~(\ref{hadnres}), it is a great advantage to have 
the parton densities and fragmentation functions in moment space. 
Technically, since the parton distributions are typically only available 
in $x$ space, we first perform a fit of a simple functional form to each 
distribution, of which we are then able to take Mellin moments analytically. 
This is done separately for each parton type and at each scale. 

As we have previously mentioned, the dependence of the spin-dependent cross
section on the azimuthal angle $\phi$ is on the form $\cos(2 \phi)$. 
As in~\cite{Muk} our convention for the GSI-FAIR and J-PARC cases 
is to integrate the transversely 
polarized cross section over the four quadrants in $\phi$ with alternating 
signs, in the form 
$(\int^{\frac{\pi}{4}}_{-\frac{\pi}{4}}-\int^{\frac{3\pi}{4}}_{\frac{\pi}{4}}
+\int^{\frac{5 \pi}{4}}_{\frac{3\pi}{4}}-
\int^{\frac{7\pi}{4}}_{\frac{5\pi}{4}}) 
\, \cos(2 \phi) d\phi = 4$. The unpolarized cross section is integrated over
all $\phi$, resulting in a factor $2\pi$. For RHIC, we tailor our
results to the PHENIX detector which covers only half of the pion's 
azimuthal angle. We therefore follow~\cite{Muk} to integrate only over 
the two quadrants $- \pi /4 < \Phi < \pi /4$ and $3\pi /4 < \Phi < 5\pi /4$
here, which gives $(\int^{\frac{\pi}{4}}_{-\frac{-\pi}{4}} +
\int^{\frac{5 \pi}{4}}_{\frac{3\pi}{4}})\, \cos(2 \phi) d\phi = 2$, and
$\pi$ for the spin-averaged cross section. It would be straightforward
to adapt our calculations to the STAR detector as well. At mid-rapidity,
we expect results very similar to the ones shown for the PHENIX case
below. It will be very worthwhile to also investigate the prospects offered 
by STAR's new capabilities~\cite{Bland:2005uu} at very forward rapidities.

\begin {figure}[t]
\begin{center}
\includegraphics[width = 6.0in]{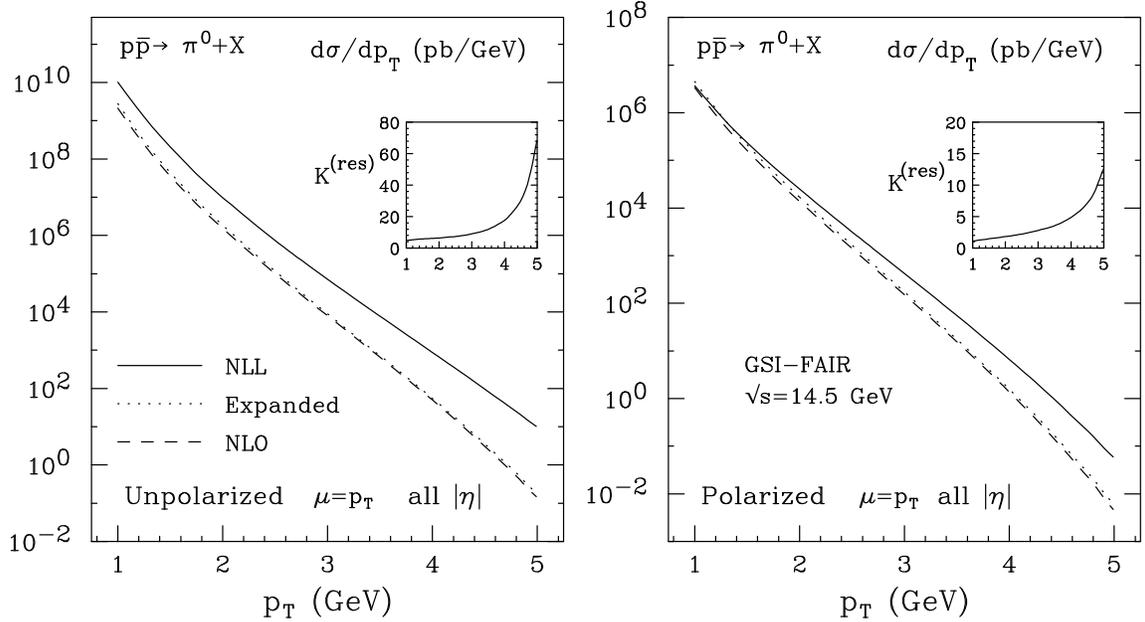}  
\end{center}
\caption {\it{Fully rapidity-integrated NLL resummed cross section, its 
expansion to ${\cal O}(\as^3)$, and the NLO cross section, for unpolarized 
(left) and transversely polarized (right) $\bar{p}p \to \pi^0 X $ at 
$\sqrt{S}=14.5$ GeV. In the insets we present the ratios between the 
NLL resummed and the NLO cross sections.}\label{fig1} }
\end{figure}

We first present our results for $\bar{p}p$ collisions 
at $\sqrt{S}=14.5$ GeV, corresponding to the GSI-FAIR project. 
At this relatively modest energy one expects the perturbative 
threshold corrections that we resum here to be particularly important. 
In Fig.~\ref{fig1} we show the rapidity-integrated spin-averaged 
(left) and spin-dependent (right) cross sections at this energy.
We display separately the NLL resummed cross section, its first-order 
(${\cal O}(\as^3)$) expansion, and the NLO one. 

As can be observed for both the unpolarized and transversely polarized
cross sections, the ${\cal O}(\as^3)$ expansion faithfully reproduces the NLO 
result, implying that higher-order corrections are indeed dominated by the
threshold logarithms. 
In the upper right corner of each figure, we show the ``K-factor'' for the 
resummed cross section over the NLO one, 
\begin{equation}
\label{eq:kres}
K^{{\rm (res)}} = \f{{d\sigma^{\rm (match)}}/{dp_T d\phi}}
{{d\sigma^{\rm (NLO)}}/{dp_T d\phi}}\, .
\end{equation}
It is interesting to see that the $K$-factors are very large, meaning that 
resummation results in a large enhancement over NLO. It is worth mentioning 
that a previous study of the DY process at this energy also found very large 
$K$-factors for the cross sections~\cite{Shim}. 

\begin {figure}[t]
\begin{center}
\includegraphics[width = 3.1in]{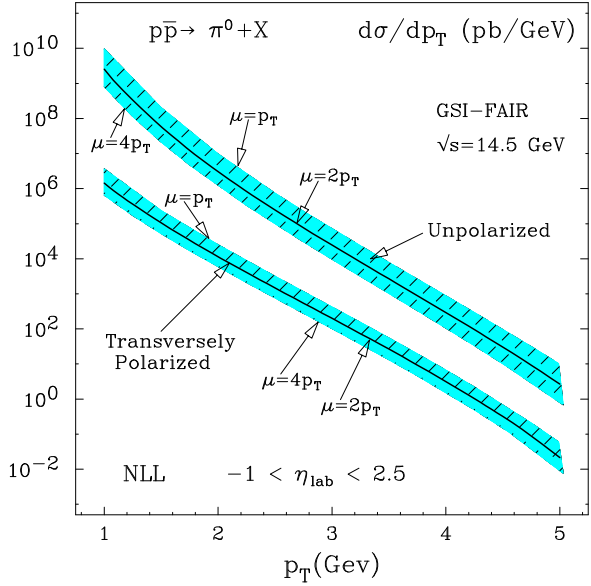}  
\end{center}
\caption {\it{ NLL resummed cross section for unpolarized and transversely 
polarized 
$\bar{p}p \to \pi^0 X $ at $\sqrt{S}=14.5$ GeV for $-1<\eta_{lab}<2.5$. 
The shaded bands represent the changes of the results if the 
factorization/renormalization scale is varied in the range 
$p_T \leq \mu \leq 4 p_T$. The solid line corresponds to $\mu=2p_T$.}
\label{fig2} }
\end{figure}

To match the experimental conditions more realistically, we have to take into 
account the rapidity range that would be covered by the experiments. 
We assume this range to be $-1<\eta_{{\mathrm{lab}}}<2.5$, where 
$\eta_{{\mathrm{lab}}}$ is the 
pseudorapidity of the pion in the laboratory frame. We count positive 
rapidity in the forward direction of the antiproton. $\eta_{{\mathrm{lab}}}$ 
is related to the c.m. pseudorapidity $\eta_{cm}$ by
\begin{equation}
\eta_{{\mathrm{lab}}}=\eta_{{\mathrm{cm}}}+\frac{1}{2}\ln 
\frac{E_{\bar{p}}}{E_{p}} \; .          
\end{equation}
Therefore the rapidity interval that we use roughly corresponds to 
$|\eta_{cm}|\lesssim 1.75$ in the c.m. system. To obtain a resummed cross
section for this interval, we use the approximation
\begin{equation}
\f{p_T^3\, d\sigma^{\rm (match)}}{dp_T d\phi}({\rm \eta\; in\, 
experimental \, range})
= K^{{\rm (res)}} \, \f{p_T^3\, d\sigma^{\rm (NLO)}}{dp_T d\phi }
({\rm \eta\; in\, experimental \, range})\, ,
\end{equation}
where $K^{{\rm (res)}}$ is as defined in Eq.~(\ref{eq:kres}) in terms of 
cross sections integrated over the full region of rapidity. In other words, 
we ``re-scale'' the matched resummed cross section by the ratio of NLO cross 
sections integrated over the experimentally relevant rapidity region 
or over all $\eta$, respectively. For the $p_T$ values we are considering 
here, the region  $-1<\eta_{{\mathrm{lab}}}<2.5$ in fact almost coincides 
with the full kinematically allowed $\eta_{cm}$ range.
The results for the NLL resummed 
cross sections integrated over the range $-1<\eta_{{\mathrm{lab}}}<2.5$ 
are shown in Fig.~\ref{fig2}. We also present in the figure the uncertainties 
in the prediction resulting from variation of  the scale $\mu$ in the range 
$p_T \leqslant \mu \leqslant 4p_T$. One can see that the scale uncertainty 
remains rather
large even after resummation.
\begin {figure}[!ht]
\begin{center}
\includegraphics[width = 3.1in]{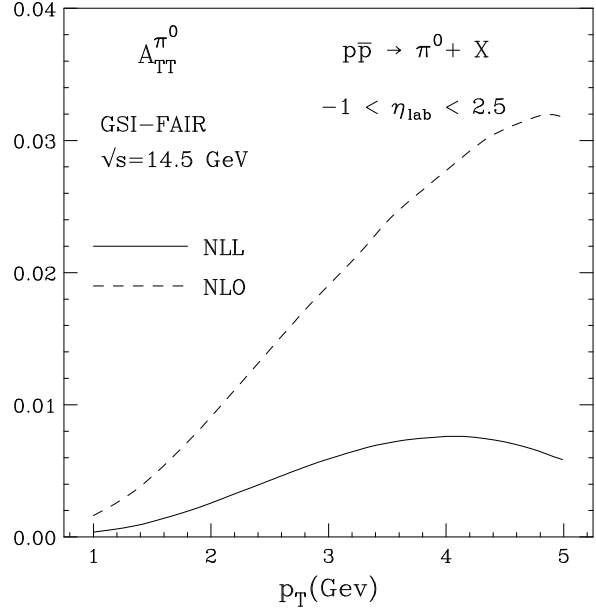}  
\end{center}
\caption {\it {NLL and NLO results for the double spin asymmetry $A^{\pi}_{TT}$
in $\bar{p}p$ collisions at $\sqrt{S}=14.5$~GeV, using ``model'' transversity
distributions that saturate the Soffer bound~\cite{Soffer} at a low scale.
}\label{fig3} }
\end{figure}

We next investigate how NLL resummation influences the double-spin asymmetry  
$A^{\pi}_{TT}$ for $\pi^0$ production. The results can be seen 
in Fig.~\ref{fig3}, where we have chosen the scale $\mu=p_T$. As one can 
observe, there is a significant 
decrease of  $A_{TT}^{\pi}$ when NLL resummation is included. Even after 
resummation the asymmetry appears to be large enough
to be accessible experimentally.

We now turn to $p^\uparrow p^\uparrow\to hX$ processes and, 
as we have mentioned previously, we will discuss neutral pion 
production at $\sqrt{S}=62.4$ GeV and at $\sqrt{S}=10$ GeV,
as relevant for experiments at RHIC (here, PHENIX) and J-PARC, 
respectively. 
\begin {figure}[!ht]
\begin{center}
\includegraphics[width = 5.92in]{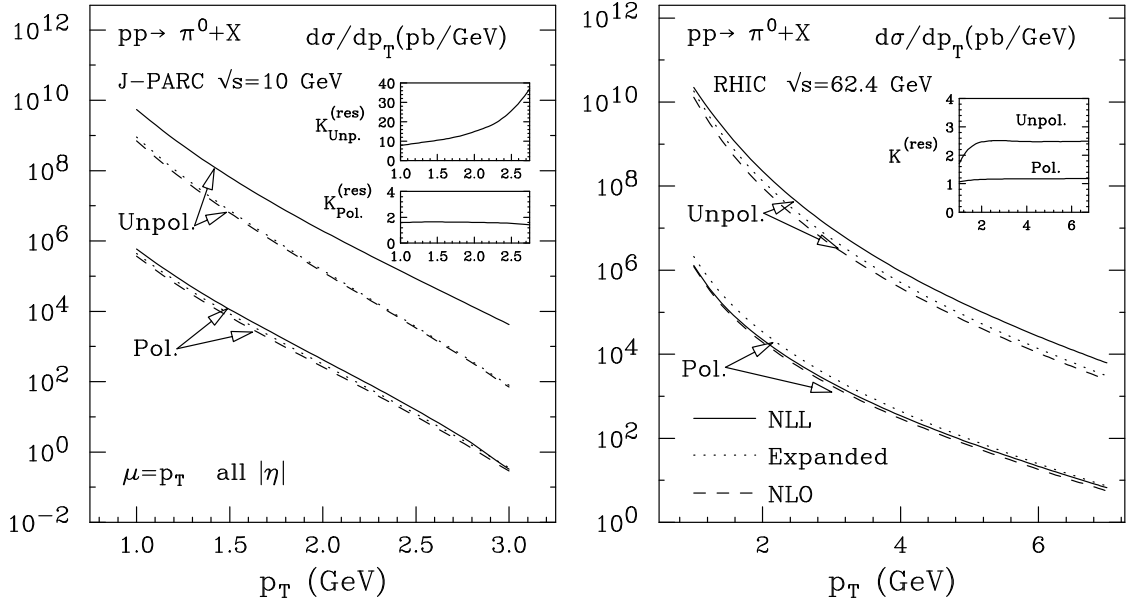}  
\end{center}
\caption {\it{Fully rapidity-integrated NLL resummed cross section, its 
expansion to ${\cal O}(\as^3)$, and the NLO cross section, for unpolarized 
and transversely polarized $pp \to \pi^0 X $ at $\sqrt{S}=10$ GeV (left), 
and at $\sqrt{S}=62.4$ GeV (right). In the insets we present the ratios 
between the NLL resummed and the NLO cross sections.}\label{fig4} }
\end{figure}

As before in Fig.~\ref{fig1} we display in Fig.~\ref{fig4} the 
rapidity-integrated spin-averaged and spin-dependent cross sections for 
J-PARC (left) and RHIC (right). Again,  the ${\cal O}(\as^3)$ expansion faithfully reproduces the NLO 
result, implying that the threshold logarithms addressed by resummation 
dominate the cross section in these kinematic regimes. This holds true,
in particular, at $\sqrt{S}=10$ GeV where one is much closer to the 
threshold regime. The agreement is more noticeable for the unpolarized cross section 
because of the large weight of the logarithmic contributions arising from the initial state gluonic 
subprocesses, whereas for transversely polarized 
scattering the expanded results slightly overestimate the NLO ones. 

As can be expected, 
the effects of resummation are much larger at $\sqrt{S}=10$ GeV than at 
$\sqrt{S}=62.4$ GeV. It is also interesting to notice that the $K$-factor 
is again much larger for the spin-averaged cross section than for the 
spin-dependent case, especially at $\sqrt{S}=10$ GeV. This immediately 
implies that the spin asymmetry $A^{\pi}_{TT}$ will be dramatically 
reduced when going from NLO to the NLL resummed case.      
    
\begin {figure}[t]
\begin{center}
\includegraphics[width = 5.9in]{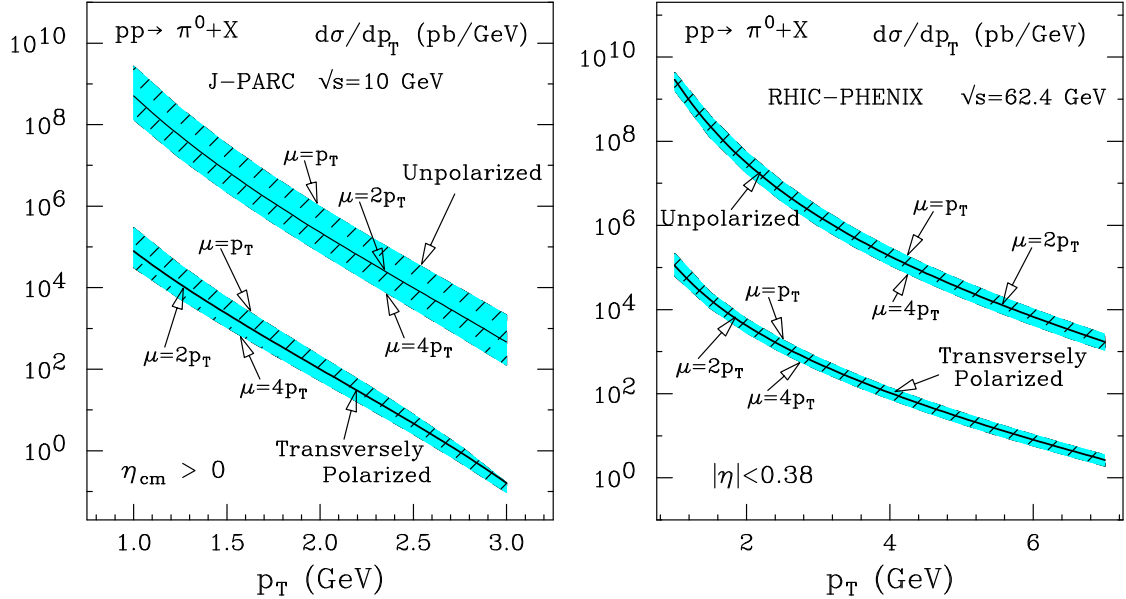}  
\end{center}
\caption {\it{ NLL resummed cross section for unpolarized and transversely 
polarized $p p \to \pi^0 X $ at fixed-target $\sqrt{S}=10$ GeV for 
$\eta_{cm} > 0$ (left) and at $62.4$ GeV for $|\eta| < 0.38$ (right).
The shaded bands represent the changes of the results if the 
factorization/renormalization scale is varied in the range 
$p_T \leq \mu \leq 4 p_T$. The solid line corresponds to $\mu=2p_T$.}
\label{fig5} }
\end{figure}

In order to allow comparison of our theoretical predictions with future 
data we consider the regions of pseudo-rapidity as 
$\eta_{cm} > 0$ and $|\eta|<0.38$ for J-PARC and RHIC (PHENIX), respectively,
over which we again integrate.
The first choice is based on the assumption of a forward spectrometer 
geometry with 200 mrad acceptance, similar to the one used by the COMPASS 
experiment at CERN, as it was also considered in Ref.~\cite{Riedl:2007sv}. 
The results for the two cases are presented in Fig.~\ref{fig5} which 
shows the NLL spin-averaged and spin-dependent cross sections. As before 
we have taken into account the theoretical uncertainties in the factorization 
and renormalization scales by varying $\mu=\zeta p_T$, with $\zeta=1,2,4$. 
It is worth noticing that after resummation the scale dependence is 
considerably reduced in the case of RHIC.

\begin {figure}[!ht]
\begin{center}
\includegraphics[width = 5.9in]{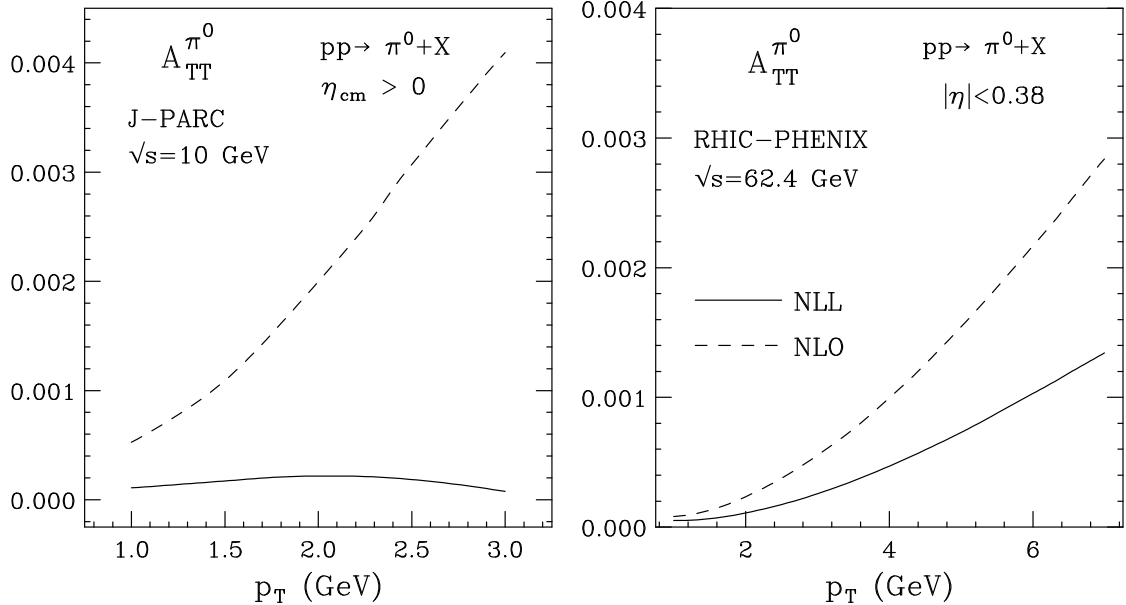}  
\end{center}
\caption {\it{ Same as Fig.~\ref{fig3}, but for $p p$ collisions at 
$\sqrt{S}=10$~GeV (left) and at $\sqrt{S}=62.4$~GeV (right).}\label{fig6} }
\end{figure}

Finally, we show in Fig.~\ref{fig6} our theoretical predictions for the 
spin-asymmetry $A_{TT}^{\pi^0}$ at both J-PARC (left) and RHIC (right).
As we anticipated, the NLL corrections significantly reduce the spin asymmetry 
with respect to the NLO results. This reduction is very significant for 
fixed-target experiments at J-PARC, whereas it is much more modest at RHIC's 
$62.4$ GeV.

\section{Conclusions  \label{sec5}}

We have studied in this paper the NLL resummation of threshold
logarithms in the partonic cross sections relevant for the processes
$pp\to h X$ and $\bar{p}p \to h X$ at high transverse momentum of the 
hadron $h$, when the initial nucleons are transversely polarized. 
We have applied the resummation to proton-antiproton scattering at 
$\sqrt{S}=14.5$ GeV, at which experiments might be carried out at 
GSI-FAIR, and to proton-proton scattering at  $\sqrt{S}=62.4$ GeV and 
at $\sqrt{S}=10$ GeV, relevant for RHIC and J-PARC, respectively. We 
find that perturbative resummation produces a large enhancement of
both the spin-averaged and the polarized cross sections. Its effect 
on the spin asymmetry is a significant reduction, especially at the 
two lower energies we have considered. We close by noting that for
these cases power-suppressed contributions to the cross sections may be 
significant as well for the kinematics we have considered and will require 
careful theoretical study in the future.

\section*{Acknowledgments}
The work of D.dF has been partially supported by Conicet,
 UBACyT and ANPCyT.
W.V.\ is grateful to the 
U.S.\ Department of Energy (contract number DE-AC02-98CH10886) for
providing the facilities essential for the completion of his work.
 The work of F.W. has been supported by UBACyT. 

\newpage

\appendix

\section*{Appendix}

\mysection{Results for the various subprocesses}

In this appendix we compile the expressions for the transverse-spin-dependent 
Born cross sections for the various partonic subprocesses, and for 
the polarized process-dependent coefficients $\delta C_{ab\to cd}^{(1)}$, 
$\delta G_{I\, ab\to cd}$ that contribute to Eq.~(\ref{eq:res}). All other
ingredients of the resummation formula coincide with their expressions in the 
spin-averaged case and may be found in Ref.~\cite{DW1}. Since the 
$\delta C_{ab\to cd}^{(1)}$  have rather lengthy expressions, we only 
give their numerical values for number of active flavors $N_f=5$, 
and for factorization and renormalization scales set to $\mu=Q=\sqrt{2}p_T$. 
In all expressions below, $C_A=3$ and $C_F=(C_A^2-1)/2C_A=4/3$. The
lower index $I$ of the coefficients $\delta G_{I ab\to cd}$ given below runs over the elements of the 
relevant color basis in each case. Also, $B(a,b)$ is the Beta-function.

\begin{description}
\item[$q\bar{q}\to q'\bar{q'}$:]
\begin{eqnarray}
\frac{ d \delta \hat{\sigma}^{{\rm (Born)}}_{q\bar{q}\to q'\bar{q'}} (N)}
{d \phi} &=& \alpha_s^2 \, \frac{\pi}{15}\frac{C_F}{C_A} \, N(N+1)(N+2) \, 
\cos(2 \phi) \, B\left(N,\frac{7}{2}\right) \ , \nn \\
\delta G^{(1)}_{1\, q\bar{q}\to q'\bar{q'}}=1 \ , && \delta C^{(1)}_{1\, 
q\bar{q}\to q'\bar{q'}}=C^{(1)}_{1\, q\bar{q}\to q'\bar{q'}} \ . 
\end{eqnarray}
\item[$qq\to qq$:]
\begin{eqnarray}
&& \frac{d \delta \hat{\sigma}^{{\rm (Born)}}_{qq\to qq} (N)}{d \phi} = 
\alpha_s^2 \, \frac{\pi}{2} \frac{C_F}{C_A^2} \,  \cos(2 \phi) \,  B\left(N+2, 
\frac{1}{2}\right) \; , \nn \\
&& \delta G_{1\, qq\to qq}=-1\,,\,\,\,\,\, \delta G_{2\, qq\to qq}=2\, ,
\,\,\,\,\,  \delta C^{(1)}_{1\, qq\to qq}=21.6034 \,\,(N_f=5)\; .
\end{eqnarray}
\item[$q\bar{q}\to q\bar{q}$:]
\begin{eqnarray}
&& \frac{d \delta \hat{\sigma}^{{\rm (Born)}}_{q\bar{q}\to q\bar{q}} 
(N)}{d \phi} =
\alpha_s^2 \, \frac{\pi}{4} \frac{C_F}{C_A^2}\,  \left[ C_A
(N+2)+ 2N+5\right] \, 
\cos(2 \phi) \, B\left(N+2, \frac{3}{2}\right) \; , \nn \\
&& \delta G_{1\,q\bar{q}\to q\bar{q}}=1,\,\,\,\,\, \delta G_{2\,
q\bar{q}\to q\bar{q}}=0 \,, \,\,\,\,\, \delta C^{(1)}_{1\, 
q\bar{q}\to q\bar{q}}=10.9783 \,\,(N_f=5)\; .
\end{eqnarray}
\item[$q\bar{q}\to gg$:]
\begin{eqnarray}
&&\frac{d \delta \hat{\sigma}^{{\rm (Born)}}_{q\bar{q}\to gg} (N)}{d \phi} = 
\alpha_s^2 \, \frac{\pi}{4} \frac{C_F}{C_A^2}\, \left[ C_A^2 (2N+6)- 
2(2N+5)\right] \, 
\cos(2 \phi) \,  B\left(N+2, \frac{3}{2}\right) \; , \nn \\
&& \delta C^{(1)}_{1\, q\bar{q}\to gg}= C^{(1)}_{1\, q\bar{q}\to gg}, 
\,\,\,\,\, \delta G_{1\, q\bar{q}\to gg}=  \frac{5}{7}, \,\,\,\,\, 
\delta G_{2\, q\bar{q}\to gg}=\frac{2}{7}, \,\,\,\,\, \delta 
C^{(1)}_{1\, q\bar{q}\to gg}= C^{(1)}_{1\, q\bar{q}\to gg} \,.
\end{eqnarray}
\end{description}

\mysection{Spin-dependent color-connected Born cross sections}

As discussed in~\cite{KOS,DW1}, in order to obtain the coefficients $\delta 
G_{I\, ab\to cd}$, one needs the color-connected Born cross sections. 
For convenience, 
we list them in this Appendix for the transversely polarized case. Our choices 
for the color bases are the same as in Ref.~\cite{KOS}. We will not repeat 
the formulas for the soft matrices $S$, the anomalous dimension matrices 
$\Gamma$ and the hard matrices $H$ for the unpolarized case, which have been 
given in~\cite{KOS,DW1}. The $S$ and $\Gamma$ are spin-independent and 
are therefore the same for the polarized processes. Only the hard 
matrices $\delta H$ for the polarized case are different. As in~\cite{KOS}, 
we will present our results for arbitrary partonic rapidity, 
even though for our actual study we only need the case $\hat{\eta}=0$. For 
each partonic reaction $a b \to c d $ we define the Mandelstam variables 
$s=(p_a + p_b)^2 =(p_c + p_d)^2 $, $t=(p_a - p_c)^2 =(p_b - p_d)^2$ and 
$u=(p_a - p_d)^2 =(p_b - p_c)^2$. Both $t$ and $u$ are functions of 
$\hat{\eta}$. In all expressions below, $N_c=3$ and $C_F=4/3$.

$q_j {\bar q_j}\rightarrow q_j {\bar q_j}$
\beqa
\delta H_{11}^{q_j {\bar q_j}\rightarrow q_j {\bar q_j}} &=& 
\alpha_s^2 \, \frac{4 C_F^2}{{N_c}^4} \, \frac{t u}{s^2}\,\cos(2 \phi)\, , 
\nonumber \\
\delta H_{12}^{q_j {\bar q_j}\rightarrow q_j {\bar q_j}} &=& 
\alpha_s^2 \, \frac{2 C_F}{N_c^3} \, \left( \frac{-2 t u}{N_c s^2} + 
\frac{u}{s}\right)  
\, \cos(2 \phi) =\delta H_{21}^{q_j {\bar q_j}\rightarrow q_j {\bar q_j}} 
\, ,\nonumber \\
\delta H_{22}^{q_j {\bar q_j}\rightarrow q_j {\bar q_j}} &=& 
\alpha_s^2 \, \frac{1}{N_c^3} \, \left( \frac{4 t u}{N_c s^2} - 
\frac{4 u}{s}\right)  \, \cos(2 \phi)
\, .
\eeqa

$q_j {\bar q}_j \rightarrow q_k {\bar q}_k $
\beq
\delta H^{q_j {\bar q_j}\rightarrow q_k {\bar q_k}} = \alpha_s^2 \,  \left[
                \begin{array}{cc}
                 C_F^2 \, h^{q_j {\bar q_j}\rightarrow q_k {\bar q_k}} & 
-C_F \, h^{q_j {\bar q_j}\rightarrow q_k {\bar q_k}}
\vspace{2mm} \\
                 -C_F \, h^{q_j {\bar q_j}\rightarrow q_k {\bar q_k}} &  
h^{q_j {\bar q_j}\rightarrow q_k {\bar q_k}}
               \end{array} \right] \, ,
\eeq

where $ h^{q_j {\bar q_j}\rightarrow q_k {\bar q_k}}= 
4 t u \cos(2 \phi)/(N_c^4 s^2)$.

$q {\bar q} \rightarrow  gg$
\beqa
\delta H_{11}^{q {\bar q} \rightarrow  gg} &=& \alpha_s^2 \, 
\frac{1}{{N_c}^4} \,\cos(2 \phi) \, , 
\nonumber \\
\delta H_{12}^{q {\bar q} \rightarrow  gg} &=&   N_c \, 
\delta H_{11}^{q {\bar q} 
\rightarrow  gg}=\delta H_{21}^{q {\bar q} \rightarrow  gg} \, ,\nonumber \\
\delta H_{22}^{q {\bar q} \rightarrow  gg} &=&  N_c^2 \, 
\delta H_{11}^{q {\bar q} 
\rightarrow  gg} \, ,\nonumber \\
\delta H_{13}^{q {\bar q} \rightarrow  gg} &=& \alpha_s^2 \, 
\frac{1}{{N_c}^3}\,  \frac{u-t}{s}\,  
\cos(2 \phi)=\delta H_{31}^{q {\bar q} \rightarrow  gg} \, , \nonumber \\
\delta H_{23}^{q {\bar q} \rightarrow  gg} &=&  N_c \, 
\delta H_{13}^{q {\bar q} 
\rightarrow  gg}=\delta H_{32}^{q {\bar q} \rightarrow  gg} \,  ,\nonumber \\
\delta H_{33}^{q {\bar q} \rightarrow  gg} &=& \alpha_s^2 \, 
\frac{1}{{N_c}^2}\,  
\frac{(t - u)^2}{s^2}\, \cos(2 \phi) \,  .
\eeqa

$q q \rightarrow q q $

This is the only quark-quark scattering process to consider.
\beqa
\delta H_{11}^{q q \rightarrow q q} &=& \alpha_s^2 \, \frac{4}{N_c^3} \, 
\cos(2 \phi) \, , 
\nonumber \\
\delta H_{12}^{q q \rightarrow q q} &=& -\alpha_s^2 \, \frac{2 C_F}{N_c^3} \, 
\cos(2 \phi) =\delta H_{21}^{q q \rightarrow q q}\, , \nonumber \\
\delta H_{22}^{q q \rightarrow q q} &=& 0 \, .
\eeqa

We finally note that the Born cross-sections in Mellin-moment space 
given in the 
previous Appendix can be obtained from the above results by
\begin{equation}
\frac{ d \delta \hat{\sigma}^{(Born)}_{ij\to kl}(N)}{d \phi}=\frac{\pi}{2} 
\int^1_0 dv \:(4v(1-v))^{N+1} \: {\mathrm{Tr}}[\delta H^{ij\to kl} 
S^{ij\to kl}] \; ,
\end{equation}
where $S^{ij\to kl}$ is the soft matrix for a given partonic process, 
to be found in~\cite{KOS}, and where in $\delta H^{ij\to kl}$ one has to 
set $u=-vs$ and $t=-(1-v)s$. The trace is in color space.


\end{document}